\documentclass[a4paper,11pt]{article}
\pdfoutput=1 

\usepackage{jheppub} 

\usepackage[T1]{fontenc} 
\usepackage{feyn}
\usepackage{feynmf} 

\usepackage{float}
\usepackage{multirow}

\usepackage{subfig}
\usepackage{amsmath}
\usepackage[dvipsnames]{xcolor}
\usepackage{hyperref}
\usepackage[normalem]{ulem}
\usepackage{xspace}
\usepackage{gensymb}

\title{$B$ meson anomalies and large $B^{+}\to K^{+}\nu\bar{\nu}$ in non-universal $U(1)^\prime$ models}
\author[a]{Peter Athron,}
\author[b]{R. Martinez,}
\author[a]{Cristian Sierra}

\affiliation[a]{Department of Physics and Institute of Theoretical Physics, Nanjing Normal University, Wenyuan Road, Nanjing, Jiangsu, 210023, China}
\affiliation[b]{Departamento de Física, Universidad Nacional de Colombia, Ciudad Universitaria, K. 45 No. 26-85, Bogotá D.C., Colombia}

\emailAdd{cristian.sierra@njnu.edu.cn}


\abstract{
In view of both the latest LHCb measurement of $R_{K^{(*)}}$ and the new $2.8\sigma$ deviation reported by Belle II on $B^{+}\to K^{+}\nu\bar{\nu}$ decays, we present a fit to the $B$ meson anomalies for various one and two dimensional hypothesis including complex Wilson coefficients. We show in a model-independent way that the generic non-universal $U(1)^{\prime}$ extensions of the SM, without flavour violation, fail to simultaneously fit those observables and corroborate that they can modify $\mathrm{BR}(B^{+}\to K^{+}\nu\bar{\nu})$ up to only a $10\%$. In view of this deficit, we propose a new way in which those models can accommodate the data at tree level by introducing lepton flavour violating couplings and non-diagonal elements of the charged lepton mixing matrix, with implications in future charged lepton flavour violation searches.}

\begin{document} 
\maketitle
\flushbottom

\section{Introduction}

A set of different discrepancies (or so called anomalies) between the theoretical predictions of the Standard Model (SM) and their corresponding experimental measurements for different flavour observables, has attracted the attention of the particle physics community for a long time now. The largest deviations  are from semi-leptonic quark level decays involving $b\to s$  transitions and the anomalous magnetic moment of the muon $a_\mu$. Here we will focus on $b$ to $s$ transitions\footnote{For a review of muon $g-2$ explanations see instead \cite{Athron:2021iuf}.}, where the long standing anomalies in $b\to s \mu^{+}\mu^{-}$ have been recently augmented by the first evidence of $B^{+}\to K^{+}\nu\bar{\nu}$ decays from the Belle II collaboration. 
Preliminary results were presented at the EPS 2023 conference\footnote{See \url{https://indico.desy.de/event/34916/contributions/146877/} and \url{https://indico.desy.de/event/34916/contributions/146877/}.}, give a measurement of $(2.4\pm 0.5^{+0.5}_{-0.4})\times 10^{-5}$ for the branching fraction of $B^{+}\to K^{+}\nu\bar{\nu}$ decays, which deviates by $2.8\sigma$ from the SM prediction \cite{Parrott:2022zte}. Combining this with previous data they find a new world average of $(1.4\pm 0.4)\times 10^{-5}$.   

Within the context of new physics (NP) effects, the most recent global fits to $b\to s \mu^{+}\mu^{-}$ data prefer a vector-like lepton universal (LU) coupling with a pull of around $6\sigma$ with respect to the SM \cite{Alguero:2023jeh}, though this was {\it before} the new $b\to s\nu\bar\nu$ measurement. For a recent discussion on the hadronic effects interpretation see \cite{Ciuchini:2022wbq}. Other flavour anomalies are the $B$ meson decays related to $b\to c \ell\nu_{\ell}$ transitions, the Cabibbo angle anomaly (CAA), leptonic $\tau$ decays of the form $\tau\to\mu\nu\nu$ and non-resonant di-electrons \cite{Crivellin:2022qcj} which we will not address here.

In general, it is difficult to find suitable NP models able to simultaneously accommodate all of the flavour observables, so using the most precise measurements as discriminators could help at excluding many of those NP scenarios. The simplest NP scenarios for explaining all the flavour anomalies are leptoquark models \cite{Dorsner:2016wpm,Buttazzo:2017ixm}, two Higgs doublet models (2HDMs) \cite{Iguro:2018qzf,Crivellin:2019dun,Athron:2021auq} and $Z'$ models \cite{Crivellin:2015mga,Allanach:2023uxz}. In this paper, we will focus on the latter class of models, namely gauged non-universal $U(1)'$ extensions of the SM, which are well motivated and have been extensively studied in the literature (for a review see e.g., \cite{Langacker:2008yv}).  One common feature to all non-universal $U(1)'$ models is the emergence of new exotic fermions as a result of requiring a theory free of chiral anomalies \cite{Ellis:2017nrp,Allanach:2018vjg}, adding then extra degrees of freedom that can alleviate most of the tensions with the experiments\footnote{One exception would be the anomalies in flavour changing charged currents related to $R_{D^{(*)}}$, which require the exchange of a charged particle at tree level. This fact points out, in the context of $Z'$ models, to the advocacy of different NP effects affecting both the neutral and charged anomalies \cite{Bause:2021prv,Greljo:2021npi}. A straightforward solution to this problem comes from extending a 2HDM with a $U(1)'$ symmetry.}.  In particular, those models are able to solve the $B$ meson anomalies \cite{Alok:2022pjb}, dark matter \cite{Martinez:2014rea,Blanco:2019hah}, the fermion mass hierarchy problem \cite{Diaz:2017eob} and the muon $g-2$ in the gauged $L_{\mu}-L_{\tau}$ model \cite{Ma:2001md}. However previous analyses of $Z'$ models found only small contributions to $B^{+}\to K^{+}\nu\bar{\nu}$ decays \cite{Buras:2014fpa,Alok:2019xub,Browder:2021hbl,Bednyakov:2023njo}.  Here we corroborate that in general at most such models can only explain $10\%$ of the new world average for $B^{+}\to K^{+}\nu\bar{\nu}$.   In view of this deficit, we propose a new way in which a lepton flavour violating (LFV) non-universal $U(1)'$ model could simultaneously fit the $B$ meson neutral anomalies and the latest Belle II measurement.


This paper is structured as follows: in section \ref{sec:Model-indep} we provide an update for the model-independent fits to the neutral anomalies with both real and complex Wilson coefficients. Then in section \ref{sec:Non-U1-models} we review the basic ingredients for a generic non-universal $U(1)'$ model and in section \ref{sec:Results} we present our results and predictions. Finally, we summarise our conclusions in section \ref{sec:conclusions}.

\section{Model independent fits}\label{sec:Model-indep}

We make use of the \textsf{flavio} package \cite{Straub:2018kue} with the same observables as in \cite{Greljo:2022jac}\footnote{Excepting for the LHCb differential branching ratio for $\Lambda_{b}\rightarrow\Lambda\mu^{+}\mu^{-}$,  which still suffers of large theoretical uncertainties in the low $q^2$ bins.} including the latest LHCb measurements for $R_{K^{(*)}}$ \cite{LHCb:2022qnv,LHCb:2022vje},
\begin{equation}
\begin{split}
    0.1<q^2<1.1 \begin{cases}
        R_K &= 0.994~^{+0.090}_{-0.082} (\mathrm{stat}) ^{+0.029}_{-0.027} (\mathrm{syst}), \\
        R_{K^{\ast}} &= 0.927~^{+0.093}_{-0.087} (\mathrm{stat}) ^{+0.036}_{-0.035}(\mathrm{syst}),
   \end{cases} \\
   1.1<q^2<6.0
   \begin{cases}
        R_K &= 0.949~^{+0.042}_{-0.041} (\mathrm{stat}) ^{+0.022}_{-0.022} (\mathrm{syst}), \\
        R_{K^{\ast}} &=1.027~^{+0.072}_{-0.068} (\mathrm{stat}) ^{+0.027}_{-0.026} (\mathrm{syst}),
   \end{cases}
\end{split}
\end{equation}
now in agreement with the SM prediction well within $1\sigma$. In addition, we also include the new Belle II measurement and the related upper limits reported by Belle in \cite{Belle:2017oht}, for a total of 180 observables. In particular, 96 of those observables are individual bins reported by LHCb \cite{LHCb:2020lmf}, CMS \cite{CMS:2017ivg} and ATLAS \cite{ATLAS:2018gqc} related to the coefficients in the angular distributions of $B^0\to K^{*0}\mu^+\mu^-$ and $B^+\to K^{*+}\mu^+\mu^-$ decays, described by the di-lepton invariant mass squared $q^2$ and three angles $\theta_K$, $\theta_{\ell}$ and $\phi$ \cite{Altmannshofer:2008dz},
{\small{}
\begin{align}
\frac{1}{d(\Gamma+\bar{\Gamma})/dq^{2}}\frac{d^{3}(\Gamma+\bar{\Gamma})}{d\cos\theta_{\ell}d\cos\theta_{K}d\phi} & =\frac{9}{32\pi}\left[\frac{3}{4}(1-F_{L})\sin^{2}\theta_{K}+F_{L}\cos^{2}\theta_{K}+\frac{1}{4}(1-F_{L})\sin^{2}\theta_{K}\cos2\theta_{\ell}\right.\nonumber \\
 & \qquad\qquad\qquad-F_{L}\cos^{2}\theta_{K}\cos2\theta_{\ell}+S_{3}\sin^{2}\theta_{K}\sin^{2}\theta_{\ell}\cos2\phi\nonumber \\
 & \qquad\qquad\qquad+S_{4}\sin2\theta_{K}\sin2\theta_{\ell}\cos\phi+S_{5}\sin2\theta_{K}\sin\theta_{\ell}\cos\phi\nonumber \\
 & \qquad\qquad\qquad+\frac{4}{3}A_{FB}\sin^{2}\theta_{K}\cos\theta_{\ell}+S_{7}\sin2\theta_{K}\sin\theta_{\ell}\sin\phi\nonumber \\
 & \qquad\qquad\qquad+S_{8}\sin2\theta_{K}\sin2\theta_{\ell}\sin\phi+S_{9}\sin^{2}\theta_{K}\sin^{2}\theta_{\ell}\sin2\phi\biggr],
\end{align}
}{\small\par}
where $A_{{\rm FB}}$ is the forward-backward asymmetry of the di-muon
system and $F_{L}$ is the fraction of longitudinal polarisation of
the $K^{*0}$ meson. $S_i$ are form factor dependent observables which can be related to the so called optimised $P'$ basis as \cite{Descotes-Genon:2013vna},
\begin{equation}
\begin{aligned}P_{1} & =\frac{2\,S_{3}}{(1-F_{{\rm L}})},\qquad P_{2}=\frac{2}{3}\frac{A_{{\rm FB}}}{(1-F_{{\rm L}})},\qquad P_{i}^{\prime}=\frac{S_{i}}{\sqrt{F_{{\rm L}}(1-F_{{\rm L}})}}.\end{aligned}
\label{eq:P1P2}
\end{equation}

We choose to work in the $S$ basis for all angular observables when possible. This is motivated by remarks in \cite{Bhom:2020lmk} pointing out that the optimised observables are clean from hadronic uncertainties only at leading order, besides being non-linearly correlated with each other and therefore, not offering major advantages for precision in phenomenological analysis.

\begin{table}
\begin{centering}
{\scriptsize{}}%
\begin{tabular}{|c|c|c|c|}
\hline 
{\scriptsize{}1D Hyp } & {\scriptsize{}Best-fit point } & {\scriptsize{}$\Delta\chi^{2}$ } & {\scriptsize{}$\textrm{Pull}_{\textrm{SM}}$}\tabularnewline
\hline 
\hline 
{\scriptsize{}$C_{9}^{\mu\mu}=C_{9}^{ee}$ } & {\scriptsize{}$-0.98\pm0.14$ } & {\scriptsize{}19.49 } & {\scriptsize{}$6.2\,\sigma$}\tabularnewline
\hline 
{\scriptsize{}$C_{9}^{\mu\mu}$ } & {\scriptsize{}$-0.50\pm0.11$ } & {\scriptsize{}11.68 } & {\scriptsize{}$4.8\,\sigma$}\tabularnewline
\hline 
{\scriptsize{}$C_{9}^{\prime\mu\mu}$ } & {\scriptsize{}$-0.03\pm0.09$ } & {\scriptsize{}0.00 } & {\scriptsize{}$0.3\,\sigma$}\tabularnewline
\hline 
{\scriptsize{}$C_{10}^{\mu\mu}$ } & {\scriptsize{}$0.19\pm0.07$ } & {\scriptsize{}3.46 } & {\scriptsize{}$2.6\,\sigma$}\tabularnewline
\hline 
{\scriptsize{}$C_{10}^{\prime\mu\mu}$ } & {\scriptsize{}$0.03\pm0.06$ } & {\scriptsize{}0.07 } & {\scriptsize{}$0.4\,\sigma$}\tabularnewline
\hline 
{\scriptsize{}$C_{9}^{\mu\mu}=C_{10}^{\mu\mu}$ } & {\scriptsize{}$-0.09\pm0.09$ } & {\scriptsize{}0.40 } & {\scriptsize{}$0.9\,\sigma$}\tabularnewline
\hline 
{\scriptsize{}$C_{9}^{\mu\mu}=-C_{10}^{\mu\mu}$ } & {\scriptsize{}$-0.17\pm0.05$ } & {\scriptsize{}7.20 } & {\scriptsize{}$3.8\,\sigma$}\tabularnewline
\hline 
\end{tabular}{\scriptsize{}$\quad$}%
\begin{tabular}{|c|c|c|c|c|}
\hline 
{\scriptsize{}2D Hyp } & {\scriptsize{}Best-fit point } & \multicolumn{1}{c|}{{\scriptsize{}$\rho$}} & {\scriptsize{}$\Delta\chi^{2}$ } & {\scriptsize{}$\textrm{Pull}_{\textrm{SM}}$}\tabularnewline
\hline 
\hline 
{\scriptsize{}$C_{9}^{\mu\mu}$ } & {\scriptsize{}$-0.97\pm0.14$ } & \multirow{2}{*}{{\scriptsize{}0.72}} & \multirow{2}{*}{{\scriptsize{}20.7}} & \multirow{2}{*}{{\scriptsize{}$6.1\,\sigma$}}\tabularnewline
\cline{1-2} \cline{2-2} 
{\scriptsize{}$C_{9}^{ee}$ } & {\scriptsize{}$-0.77\pm0.18$ } &  &  & \tabularnewline
\hline 
\multicolumn{1}{|c}{} & \multicolumn{1}{c}{} & \multicolumn{1}{c}{} & \multicolumn{1}{c|}{} & \tabularnewline
\hline 
{\scriptsize{}$C_{9}^{\mu\mu}$ } & {\scriptsize{}$-0.58\pm0.13$ } & \multirow{2}{*}{{\scriptsize{}-0.61}} & \multirow{2}{*}{{\scriptsize{}12.0}} & \multirow{2}{*}{{\scriptsize{}$4.5\,\sigma$}}\tabularnewline
\cline{1-2} \cline{2-2} 
{\scriptsize{}$C_{10}^{\mu\mu}$ } & {\scriptsize{}$-0.08\pm0.09$ } &  &  & \tabularnewline
\hline 
\multicolumn{1}{|c}{} & \multicolumn{1}{c}{} & \multicolumn{1}{c}{} & \multicolumn{1}{c|}{} & \tabularnewline
\hline 
{\scriptsize{}$C_{9}^{\mu\mu}$ } & {\scriptsize{}$-0.52\pm0.11$ } & \multirow{2}{*}{{\scriptsize{}0.0}} & \multirow{2}{*}{{\scriptsize{}12.2}} & \multirow{2}{*}{{\scriptsize{}$4.6\,\sigma$}}\tabularnewline
\cline{1-2} \cline{2-2} 
{\scriptsize{}$C_{9}^{ee}=C_{10}^{ee}$ } & {\scriptsize{}$0.29\pm0.25$ } &  &  & \tabularnewline
\hline 
\multicolumn{1}{|c}{} & \multicolumn{1}{c}{} & \multicolumn{1}{c}{} & \multicolumn{1}{c|}{} & \tabularnewline
\hline 
{\scriptsize{}$C_{9}^{\mu\mu}$ } & {\scriptsize{}$-0.97\pm0.14$ } & \multirow{2}{*}{{\scriptsize{}0.72}} & \multirow{2}{*}{{\scriptsize{}20.6}} & \multirow{2}{*}{{\scriptsize{}$6.1\,\sigma$}}\tabularnewline
\cline{1-2} \cline{2-2} 
{\scriptsize{}$C_{9}^{ee}=-C_{10}^{ee}$ } & {\scriptsize{}$-0.34\pm0.07$ } &  &  & \tabularnewline
\hline 
\multicolumn{1}{|c}{} & \multicolumn{1}{c}{} & \multicolumn{1}{c}{} & \multicolumn{1}{c|}{} & \tabularnewline
\hline 
{\scriptsize{}$C_{9}^{U}=C_{9}^{ee}$ } & {\scriptsize{}$-0.89\pm0.15$ } & \multirow{2}{*}{{\scriptsize{}-0.29}} & \multirow{2}{*}{{\scriptsize{}21.2}} & \multirow{2}{*}{{\scriptsize{}$6.2\,\sigma$}}\tabularnewline
\cline{1-2} \cline{2-2} 
{\scriptsize{}$C_{9}^{V}=-C_{10}^{V}$ } & {\scriptsize{}$-0.09\pm0.04$ } &  &  & \tabularnewline
\hline 
\end{tabular}{\scriptsize\par}
\par\end{centering}
\caption{Left: Common one dimensional hypothesis for real NP Wilson coefficients.
Right: Two dimensional hypothesis for real Wilson coefficients with
correlation $\rho$. The notation for the last row follows from \cite{Alguero:2018nvb}
where the short-distance Wilson coefficients can contain two types
of NP contribution, $C_{9}^{\ell\ell}=C_{9}^{U}+C_{9}^{V}$ where
the first contribution is lepton universal and the second one is lepton
flavour violating.\label{tab:1D_and_2D_cases}}
\end{table}

To simplify the analysis, none of the low bins $q^2<1.0\,\textrm{GeV}^2$ sensitive to both electromagnetic and chromomagnetic operators in the $b\to s\ell^+\ell^-$ angular observables and branching ratios were included in the fit. Besides that, and from the latest CMS-ATLAS-LHCb combination of $\mathrm{BR}(B_{s}\rightarrow\mu^{+}\mu^{-})$ presented in \cite{Greljo:2022jac}, scalar operators are strongly constrained and are taken here to be zero, allowing NP effects to come only from vector and axial-like operators. Furthermore, couplings to right-handed quarks are assumed to be negligibly small compared to the left-handed ones and therefore, the prime operators ($P_L\leftrightarrow P_R$) will not be taken into account.

In this way, the effective Hamiltonian responsible for $b\to s\ell^+\ell^-$ transitions can be written as
\begin{align}
{\cal H}_{\mathrm{eff}}^{bs\ell\ell} & =-\frac{4G_{F}}{\sqrt{2}}V_{tb}V_{ts}^{*}\sum_{i=9,10} C^{\ell\ell}_{i}\mathcal{O}^{\ell\ell}_{i},\label{eq:heff}
\end{align}
where  $\ell=e,\mu$,  $G_F$ is the Fermi constant, $V_{tb(s)}$ are elements of the Cabibbo-Kobayashi-Maskawa (CKM) matrix and the $C^{\ell\ell}_{i}$ are the effective Wilson coefficients (WCs) at the scale of the bottom quark mass, and 
\begin{align}
\mathcal{O}^{\ell\ell}_{9} & =\frac{e^{2}}{16\pi^{2}}(\bar{s}\gamma_{\mu}P_{L}b)(\bar{\ell}\gamma^{\mu}\ell),\qquad\qquad\mathcal{O}^{\ell\ell}_{10}=\frac{e^{2}}{16\pi^{2}}(\bar{s}\gamma_{\mu}P_{L}b)(\bar{\ell}\gamma^{\mu}\gamma_{5}\ell),\label{eq:basisA}
\end{align}
are the flavour-changing neutral currents (FCNC) local operators encoding the low-energy description of the high energy physics that has been integrated out. The WCs can be written as $C_{i}^{\ell\ell}=C_{i}^{\mathrm{SM}}+C_{i}^{\ell\ell,\mathrm{NP}}$ where $C_{i}^{\mathrm{SM}}$ is the SM contribution to the $i$th WC and $C_{i}^{\ell\ell,\mathrm{NP}}$ is the NP contribution. For simplicity of notation, from now on we will drop the NP index unless specified otherwise.

Here, we first review and corroborate the most common 1D and 2D hypothesis presented in \cite{Alguero:2023jeh,Greljo:2022jac,Guadagnoli:2023ddc} as well as some new 2D scenarios for the WCs from $b\to s \ell^{+}\ell^{-}$ transitions. In Table \ref{tab:1D_and_2D_cases} we provide the best fit values with $1\sigma$ uncertainties and correlations for both the 1D and 2D cases, with the
$\textrm{Pull}_{\textrm{SM}}$ metric calculated as in \cite{Capdevila:2017bsm,Capdevila:2018jhy},
\begin{align}
\textrm{Pull}_{\textrm{SM}}=\sqrt{2}\,\text{Erf}^{-1}\left[F(\Delta\chi^2;n_\text{dof})\right],
\label{eq:PullSM}
\end{align}
where $F$ is the $\chi^2$ cumulative distribution function and $n_\text{dof}$ is the number of degrees of freedom.

\begin{figure}[ht]
\begin{centering}
\includegraphics[scale=0.56]{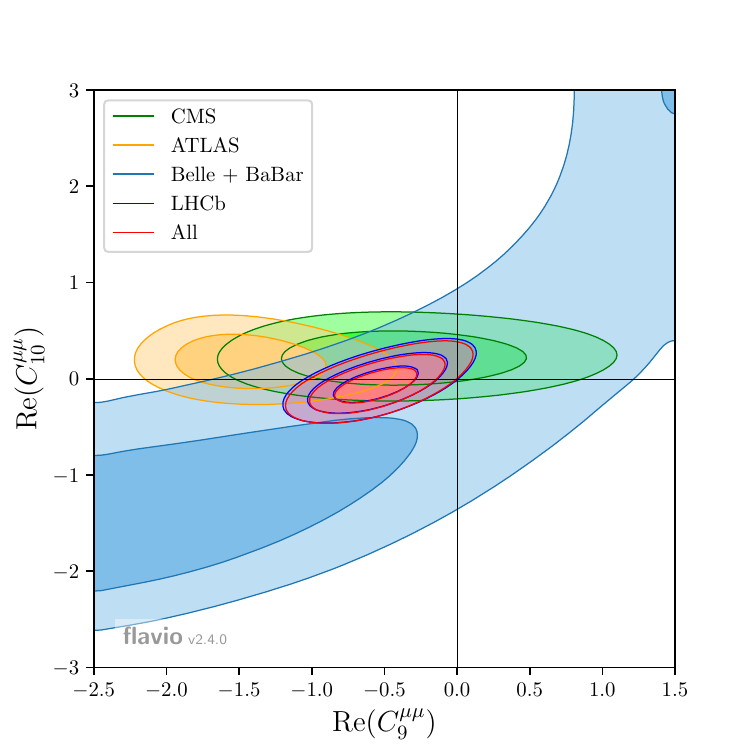} $\qquad$\includegraphics[scale=0.56]{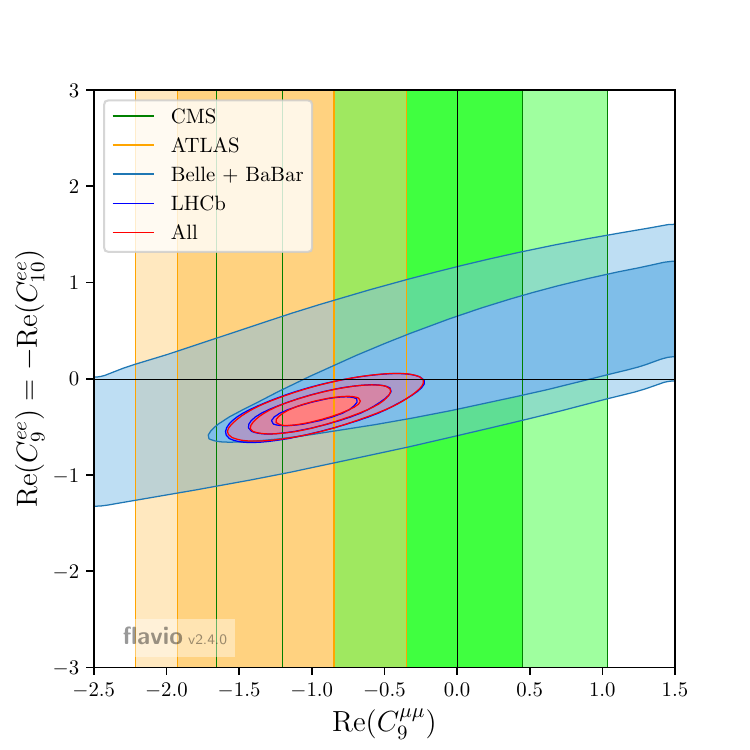} 
\par\end{centering}
\caption{Global fits showing $1\sigma$ (lighter colours) and $2\sigma$ (darker colours) regions of the $b\to s \ell^{+}\ell^{-}$ data available from LHCb (bright blue), CMS (green), ATLAS (orange), Belle and BaBar (both combined and displayed in blue). Given that the vast majority of observables come from LHCb, this experiment drives the total fit (red), almost indistinguishable from the LHCb data alone. Left: $\{C_{9}^{\mu\mu},\:C_{10}^{\mu\mu}\}$ scenario now disfavoured with respect to the scenarios fitted by $C_{9}^{\mu\mu}$ and $C_{9}^{ee}$. Right: New physics 2D hypothesis involving electronic Wilson coefficients.\label{fig:2D_WCs}}
\end{figure}

We see in Table \ref{tab:1D_and_2D_cases} that one of the most common hypothesis, namely the $\{C_{9}^{\mu\mu},\:C_{10}^{\mu\mu}\}$ 2D hypothesis, is now disfavoured by $1.6\sigma$ with respect to the scenarios fitted by $C_{9}^{\mu\mu}$ and $C_{9}^{ee}$, in accordance to the updated LHCb measurements for $R_{K^{(*)}}$ \cite{LHCb:2022qnv,LHCb:2022vje}, now SM-like\footnote{This does not mean that the $\{C_{9}^{\mu\mu},\:C_{10}^{\mu\mu}\}$ 2D hypothesis is disfavoured with respect to the SM hypothesis, giving  $\textrm{Pull}_{\textrm{SM}}=4.5\sigma$ as seen on the second row of the right table \ref{tab:1D_and_2D_cases}. For more details on this see \cite{Allanach:2022iod}.}. Particularly interesting is the new 2D scenario $\{C_{9}^{\mu\mu},\:C_{9}^{ee}=-C_{10}^{ee}\}$ (Fig.\ref{fig:2D_WCs}) with a pull of $6.1\sigma$ with respect to the SM. \\

\begin{figure}[ht]
\begin{centering}
\includegraphics[scale=0.56]{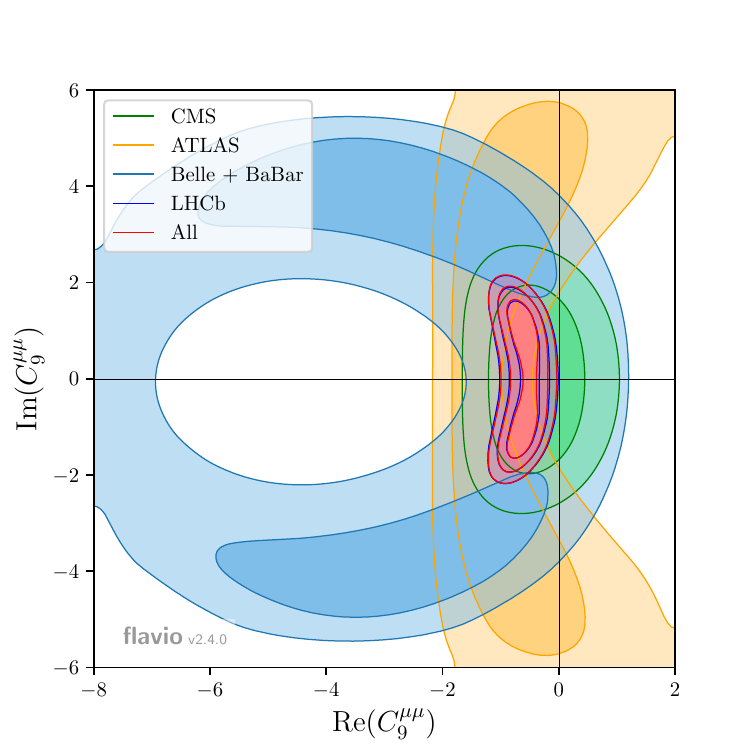} $\qquad$\includegraphics[scale=0.56]{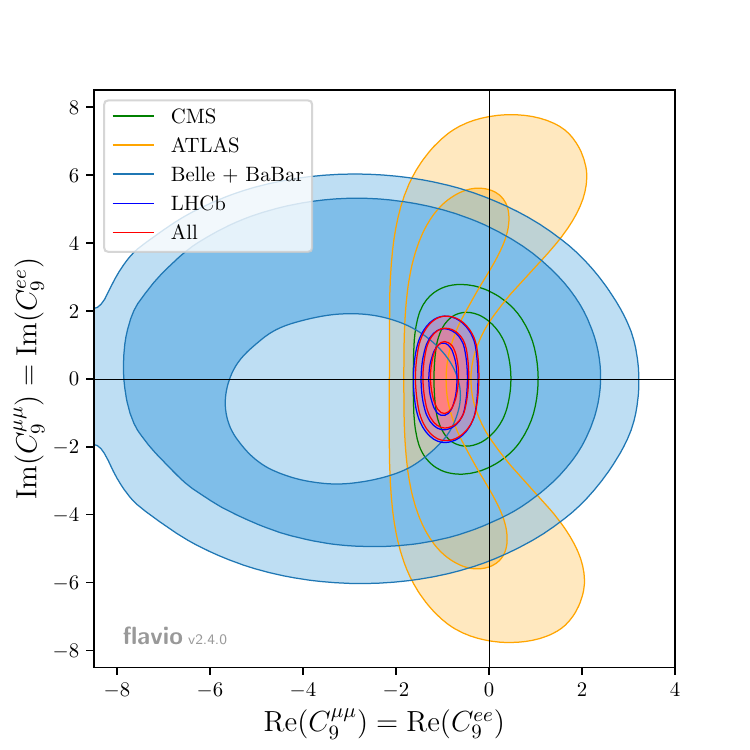} 
\par\end{centering}
\caption{Fits to the $B$ meson anomalies with complex Wilson coefficients. Left: $\{\textrm{Re}(C_{9}^{\mu\mu}),\:\textrm{Im}(C_{9}^{\mu\mu})\}$ scenario with $\Delta\chi^{2}=13.5$ and $\textrm{Pull}_{\textrm{SM}}=4.8\sigma$. Right: $\{\textrm{Re}(C_{9}^{\mu\mu})=\textrm{Re}(C_{9}^{ee}),\:\textrm{Im}(C_{9}^{\mu\mu})=\textrm{Im}(C_{9}^{ee})\}$ new case with $\Delta\chi^{2}=18.7$ and $\textrm{Pull}_{\textrm{SM}}=5.8\sigma$. Colours meaning are the same as in Fig.\ref{fig:2D_WCs}.\label{fig:complex_WCs}}
\end{figure}

\noindent In Fig.\ref{fig:complex_WCs} we also present an update for the common 2D case $\{\textrm{Re}(C_{9}^{\mu\mu}),\:\textrm{Im}(C_{9}^{\mu\mu})\}$ of complex WCs  and compare it to the new scenario $\{\textrm{Re}(C_{9}^{\mu\mu})=\textrm{Re}(C_{9}^{ee}),\:\textrm{Im}(C_{9}^{\mu\mu})=\textrm{Im}(C_{9}^{ee})\}$ studied in \cite{Guadagnoli:2023ddc}. We corroborate the findings of \cite{Altmannshofer:2021qrr,Biswas:2020uaq,Guadagnoli:2023ddc} and observe no improvement over the real WCs case, although we can see a slight preference from the Belle+BaBar data for a large negative imaginary part of $C_{9}^{\mu\mu}$.

Regarding $b\to s \nu\bar{\nu}$, if no right-handed neutrinos are included, or equivalently, if their mixing with active neutrinos is suppressed, the effective Hamiltonian at the bottom quark scale is defined as 
\begin{equation}
{\cal H}_{\mathrm{eff}}^{bs\nu_{i}\nu_{j}} = - \frac{4\,G_F}{\sqrt{2}} V_{tb}V_{ts}^*(C^{\mathrm{SM}}_{LL}\delta_{ij}+C^{\nu_{i}\nu_{j}}_{LL})\mathcal[{O}^{V}]^{\nu_{i}\nu_{j}}_{LL} + \rm h.c.,
\end{equation}
where 
\begin{equation}
C_{LL}^{\mathrm{SM}}=-X(x_{t})/\sin^{2}\theta_{w},\quad\mathrm{and}\quad\mathcal[{O}^{V}]^{\nu_{i}\nu_{j}}_{LL}=\frac{e^{2}}{16\pi^{2}}(\bar{s}\gamma_{\mu}P_{L}b)(\bar{\nu}_{i}\gamma^{\mu}(1-\gamma_{5})\nu_{j}),
\end{equation}
with $x_t=m_t^2/m_W^2$ and the function $X(x_t)$ is defined at the next-to-leading order in QCD in \cite{Misiak:1999yg,Buchalla:1998ba}, giving $C^{\mathrm{SM}}_{LL} = - 6.38 \pm 0.06$ \cite{Altmannshofer:2009ma}. In this way, we will have at first 9 WCs coming from the purely left-handed $\mathcal[{O}^{V}]^{\nu_{\ell}\nu_{\ell'}}_{LL}=\mathcal{O}^{\ell\ell'}_9-\mathcal{O}^{\ell\ell'}_{10}$ operators contributing to the branching ratio \cite{Browder:2021hbl},
\begin{align}
\label{eq:KBR}
\mathrm{BR}(B^+\to K^+ \nu\bar\nu) = \mathrm{BR}(B^{+}\to K^{+}\nu\bar\nu)_{\mathrm{SM}} \times & \Bigg[ \frac{1}{3}\,\sum_{i=1}^3  \bigg| 1 +\frac{C_{LL}^{\nu_i\nu_i}}{C_{LL}^\text{SM}}\bigg|^2 + \sum_{i\neq j} \bigg| \frac{C_{LL}^{\nu_i\nu_j}}{C_{LL}^\text{SM}} \bigg|^2\Bigg]\,,
\end{align}
with the SM prediction given by $\mathrm{BR}(B^{+}\to K^{+}\nu\bar\nu)_{\mathrm{SM}}\simeq4.4\times 10^{-6}$ from the \textsf{flavio} package.

In the most general case for real WCs, we will have four independent flavour conserving coefficients associated to the $\mathcal{O}^{\ell\ell}_9$ and $\mathcal{O}^{\ell\ell}_{10}$ operators, summing up to a total of 13 WCs for the fit. In Table \ref{tab:5D fit} we can see again the well-known preference for the muonic WC $C_{9}^{\mu\mu}\simeq-1.0$ accompanied in this case by a small positive $C_{10}^{\mu\mu}$ and our data shows a new interesting correlation $C_{9}^{ee}\simeq-C_{10}^{ee}=-0.5$ for the electronic WCs. As for the Belle II measurement, it gets fitted by a LU violating coupling with di-electron neutrinos equal to $C_{LL}^{\nu_{e}\nu_{e}}\simeq-7$ with order $7\%$ correlations with the muonic WCs. Taking these last values and correlations into account, we proceed now to map the WCs into the parameter space for a generic non-universal $Z^{\prime}$ model.

\begin{table}
\begin{centering}
\begin{tabular}{|c|c|c|c|c|c|c|}
\hline 
WC & Best-fit point & \multicolumn{5}{c|}{$\rho$}\tabularnewline
\hline 
\hline 
$C_{9}^{\mu\mu}$ & $-0.97\pm0.14$ & 1.00 & -0.06  & 0.29 & -0.02 & -0.05\tabularnewline
\hline 
$C_{10}^{\mu\mu}$ & $0.20\pm0.11$ &  & 1.00 & -0.08  & 0.27  & 0.07\tabularnewline
\hline 
$C_{9}^{ee}$ & $-0.5\pm0.5$ &  &  & 1.00 & 0.89  & -0.01\tabularnewline
\hline 
$C_{10}^{ee}$ & $0.5\pm0.4$ &  &  &  & 1.00 & 0.02\tabularnewline
\hline 
$C_{LL}^{\nu_{e}\nu_{e}}$ & $-6.8\pm1.7$ &  &  &  &  & 1.00\tabularnewline
\hline 
\end{tabular}
\par\end{centering}
\caption{Results from a 13-dimensional fit to the \textbf{$B$} meson anomalies and the Belle
II measurement. The remaining 8 WCs associated to $b\to s \nu\bar{\nu}$ decays have very large uncertainties compatible with zero and we do not display them. Here $\Delta\chi^{2}=22.9$ and $\textrm{Pull}_{\textrm{SM}}=5.9\,\sigma$.\label{tab:5D fit}}
\end{table}

\section{Non-universal $U(1)^{\prime}$  models}\label{sec:Non-U1-models}

These kinds of models assume that the dominant NP contribution to $b\to s \ell^{+}\ell^{-}$ and $b\to s \nu\overline{\nu}$ transitions comes from tree level exchange of a $SU(2)_L$ singlet $Z^{\prime}$ gauge boson associated to an extra $U(1)^\prime$ symmetry \cite{Buras:2014fpa}. This symmetry is broken spontaneously at the TeV scale by the vacuum expectation value (VEV) of a new neutral scalar $\chi$, making the $Z^{\prime}$ boson to acquire a mass term $M_{Z'}=g_X\upsilon_\chi$ with $g_X$ the new gauge coupling. The $Z^{\prime}$ will also mix with the SM $Z$ boson in the kinetic terms of the Lagrangian, however we will consider such mixing negligible\footnote{Gauge kinetic mixing was investigated in \cite{Buras:2021btx}, but the fits found this to be consistent with no kinetic mixing.}. The purely new physics interaction Lagrangian with fermions in the flavour basis is defined then by,
\begin{equation}
\mathcal{L}_{Z'f\bar{f}}=\sum_{i,j,f_{L}}g_{L}^{ij}\overline{f}_{L}^{i}\gamma^{\mu}P_{L}f_{L}^{j}Z_{\mu}^{\prime}+\sum_{i,j,f_{R}}g_{R}^{ij}\overline{f}_{R}^{i}\gamma^{\mu}P_{R}f_{R}^{j}Z_{\mu}^{\prime},
\end{equation}
where, under the same assumptions as before, the relevant couplings will be $g_{L}^{bs},$ $g_{L,R}^{\ell\ell}$ and $g_{L}^{\nu\nu}$, and the WCs generated at tree level by the effective Hamiltonian in Eq.(\ref{eq:heff}) are given by,
\begin{align}
C_{9}^{\ell\ell} & =-\frac{\pi}{\sqrt{2}G_{F}\alpha V_{tb}V_{ts}^{*}}\frac{g_{L}^{bs}(g_{L}^{\ell\ell}+g_{R}^{\ell\ell})}{M_{Z'}^{2}},\label{eq:C9ll}\\
C_{10}^{\ell\ell} & =\frac{\pi}{\sqrt{2}G_{F}\alpha V_{tb}V_{ts}^{*}}\frac{g_{L}^{bs}(g_{L}^{\ell\ell}-g_{R}^{\ell\ell})}{M_{Z'}^{2}},\label{eq:C10ll}
\end{align} 

\noindent where the lepton couplings are defined as, 
\begin{equation}
g_{L,R}^{\ell\ell}=V_{L,R}^{l\dagger}X_{L,R}V_{L,R}^{l},\label{eq:gLR}
\end{equation}
and can be in general non-universal, determined by both the $U(1)'$ charges 

\begin{equation}
X_{L,R}=\mathrm{diag}\{X_{L,R}^{e},\,X_{L,R}^{\mu},\,X_{L,R}^{\tau}\},   
\end{equation}
and by the $n\times n$ rotation matrix $V_{L}^{l}$ which transforms flavour eigenstates to mass eigenstates with $n=3+E$, where $E$ is the number of exotic charged leptons. For concreteness, we define the SM part of $V_{L}^{l}$ as
\begin{equation}
V_{L}^{l}{}^{\mathrm{SM}}=\left(\begin{array}{ccc}
1 & \sin\delta & \sin\epsilon\\
-\sin\delta & \cos\alpha_{L} & \sin\alpha_{L}\\
-\sin\epsilon & -\sin\alpha_{L} & \cos\alpha_{L}
\end{array}\right),\label{eq:VLe}
\end{equation}
where $\alpha_L$, $\delta$ and $\epsilon$ are free parameters of the rotation matrix constrained by both unitarity and the masses of the charged leptons. This texture is justified by recent searches of lepton flavour violation in Higgs boson decays from both ATLAS \cite{ATLAS:2023mvd} and CMS \cite{CMS:2016cvq,CMS:2021rsq}, reporting upper limits on the branching ratios of $H\to\tau\mu$, $H\to\tau e$ and $H\to\mu e$ decays, consistent with decoupling of 1-2 and 1-3 rotations. On the other hand, we will take the matrix $V_{R}^{l}$ to be diagonal,
i.e., with lepton universal right-handed couplings, $X_{R}=X_{R}^{e}=X_{R}^{\mu}=X_{R}^{\tau}$.\\

Regarding the $b\to s\nu\overline{\nu}$ process, the WC in question according to Table \ref{tab:5D fit}
is given by,
\begin{equation}
C_{LL}^{\nu_{e}\nu_{e}}=-\frac{\pi}{\sqrt{2}G_{F}\alpha V_{tb}V_{ts}^{*}}\frac{g_{L}^{bs}g_{L}^{\nu_{e}\nu_{e}}}{M_{Z'}^{2}},\label{eq:CLL}
\end{equation}
with $g_{L}^{\nu_{e}\nu_{e}}=(V_{L}^{\nu_{l}\dagger})_{1k}X_{L}(V_{L}^{\nu_{l}})_{k1}$
from the $SU(2)_{L}$ symmetry and $V_{L}^{\nu_{l}}=V_{L}^{l}{}^{\mathrm{SM}}V_{\mathrm{PMNS}}$ with $V_{\mathrm{PMNS}}$ the Pontekorvo-Maki-Nakagawa-Sakata (PMNS) matrix, meaning that depending on the ordering (inverted or normal) for the neutrino mixing parameters, the lepton couplings entering in Eq.(\ref{eq:CLL}) may differ. 

\section{Results}\label{sec:Results}

Equipped with the required WCs both from the model-independent fit and for the $U(1)'$ models, we define the quadratic approximation to the likelihood function 
\begin{equation}
\log\mathcal{L}=-\frac{\chi^{2}}{2},\quad\chi^{2}(\mathbf{C})\approx\chi_{min}^{2}+\frac{1}{2}\left(\mathbf{C}-\mathbf{C}_{\mathrm{bf}}\right)^{T}\mathrm{Cov}^{-1}\left(\mathbf{C}-\mathbf{C}_{\mathrm{bf}}\right),
\end{equation}
where $\mathbf{C}=\{C_{9}^{\mu\mu},\:C_{10}^{\mu\mu},\:C_{9}^{ee},\:C_{10}^{ee},\:C_{LL}^{\nu_{e}\nu_{e}}\}$ is defined in terms of the WCs in Eqs.(\ref{eq:C9ll}-\ref{eq:CLL}), $\mathbf{C}_{\mathrm{bf}}$ is given by the central values in the second column of Table \ref{tab:5D fit}, and $\mathrm{Cov}$ is the covariance matrix or Hessian associated to the correlation matrix $\rho$ also in Table \ref{tab:5D fit} obtained by using the \texttt{MIGRAD} minimization algorithm. With this function at hand, a random generator in \texttt{Mathematica} is requested to find points inside the ellipsoids defined by 
$\Delta\chi^{2}\leq\sigma, 2\sigma$ for 2 degrees of freedom and boundaries defined for two scenarios: 
\begin{enumerate}
\item Scenario with lepton flavour universality violation (LFUV).
\item  Scenario with lepton flavour violation (LFV) in addition to LFUV.
\end{enumerate}
For the first scenario we scan over the following couplings
\begin{equation}
g_{L,R}^{\mu\mu}\in[-2,\,2],\qquad g_{L}^{bs},\,g_{L,R}^{ee}\in[-0.2,\,0.2],\qquad M_{Z'}\in[1,\,10]\,\mathrm{TeV},
\end{equation}
where the ranges are chosen based on the results from \cite{Buras:2021btx}, where a global analysis was performed constraining the lepton couplings from various leptonic processes mediated at tree level by a generic $Z'$ and the muonic (electronic) couplings were found to be at most of order 2 (0.2). Regarding the values for the  $Z'$ mass, we follow \cite{Allanach:2019mfl} and \cite{DiLuzio:2019jyq} and set the lower and upper mass values for the scan as 1 and 10 TeV respectively. We also evade more stringent constraints from colliders given that, as in \cite{Buras:2014fpa}, we do not specify the flavour-conserving couplings to first and second generation quarks and therefore the high-mass Drell-Yan bounds will be suppressed assuming that the  $Z'$ boson couples dominantly to bottom quarks \cite{Greljo:2022jac} and to leptons \cite{Buras:2021btx}.

For the LFV scenario, we additionally scan over the following parameters,
\begin{equation}
X_{L}^i, X_R\in[-\sqrt{4\pi},\,\sqrt{4\pi}],\qquad\alpha_{L}\in[0,\,2\pi],\qquad\delta,\,\epsilon\in[0,\,0.5],\end{equation}
for $i=e,\mu,\tau$, $X_R$ is the single flavour universal charge for the right-handed leptons and the flavour non-universal charges for the left-handed lepton couplings are related to the lepton couplings by Eq.(\ref{eq:gLR}). As a consequence of lepton flavour violation in this scenario, the couplings of Eq.(\ref{eq:gLR}) will be non-diagonal allowing for processes as $\mu\to e\gamma$, which we take explicitly into account as a constraint from the latest measurement made by the MEG collaboration \cite{MEG:2016leq},

\begin{equation}
\textrm{BR}(\mu\to e \gamma)_{\textrm{exp}}=4.2\times10^{-13}\,,\quad 90\%\,\textrm{C.L.},\label{eq:MEG_bound}
\end{equation}

with the theory prediction given by \cite{Buras:2021btx},

\begin{align}
\mathrm{BR}(\mu \to e\gamma)_{\textrm{th}}=\frac{m^3_{\mu}}{4\pi  \Gamma_\mu}\left(|c_L^{12}|^2 +|c_R^{12}|^2\right),\label{eq:BRllpgamma}
\end{align}
where $\Gamma_\mu$ is the decay width of the muon and
\begin{align}
c_L^{12}=&\frac{e}{48\pi^2 M_{Z^{\prime}}^2}\sum_{k}
\bigg(m_{\mu} \, g_{1k}^R \, g_{k2}^R -3\, m_k \, g_{1k}^R\,  g_{k2}^L + m_{e}\, g_{1k}^L \, g_{k2}^L \bigg)\,,
\label{eq:cLij}
\end{align}
with $c_R^{12}$ obtained from $c_L^{12}$ by interchanging $L$ and $R$.

\begin{figure}[ht]
\begin{centering}
\includegraphics[scale=0.35]{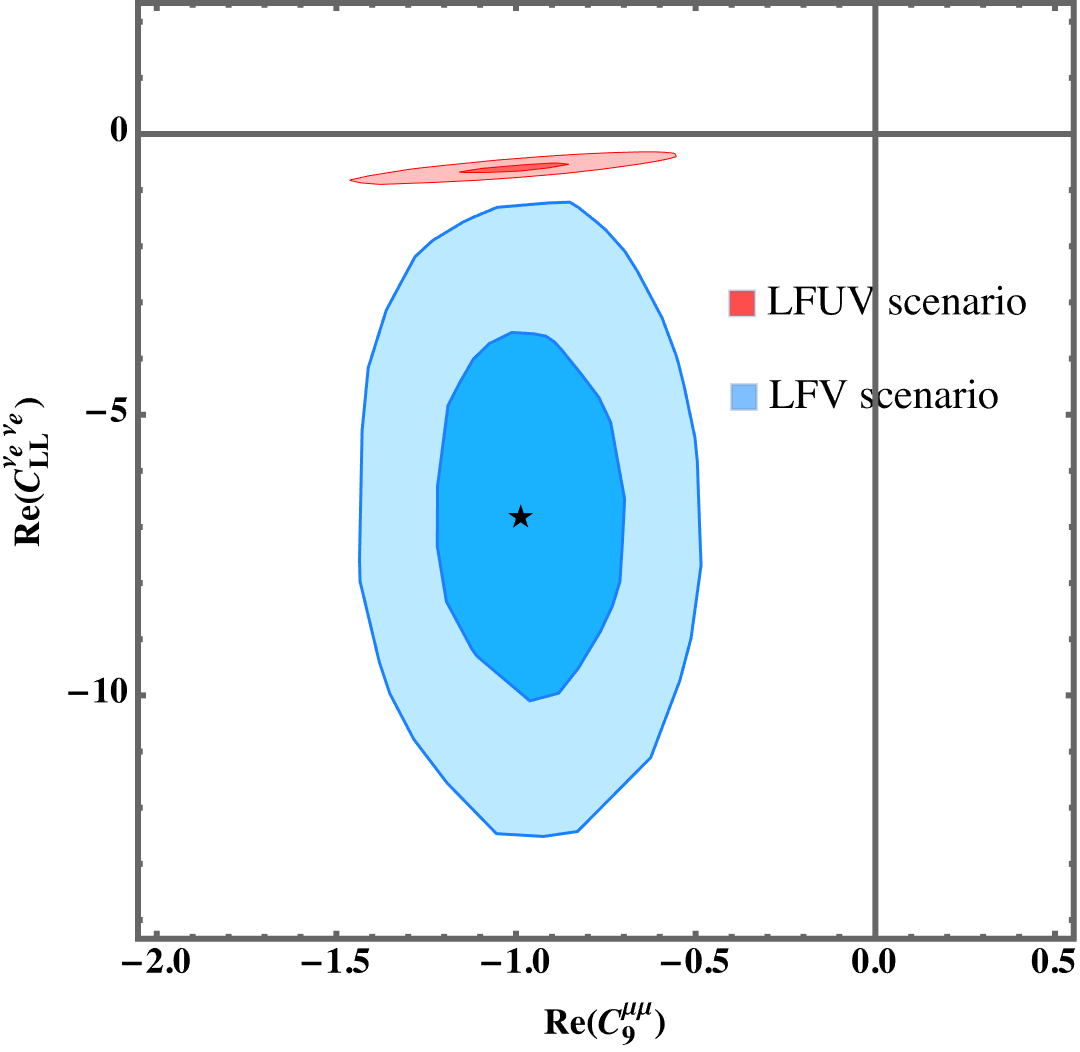} 
\par\end{centering}
\caption{Projected $1$ and $2\sigma$ regions in the $\{C_{9}^{\mu\mu},\:C_{LL}^{\nu_e\nu_e}\}$ plane fitting the $b\to s \ell^{+}\ell^{-}$ data simultaneously with $\mathrm{BR}(B^{+}\to K^{+}\nu\bar{\nu})$ at $1\sigma$ level for the LFUV scenario (red) and the LFV scenario (blue).\label{fig:fit_2_models}}
\end{figure}

We can see in Fig.\ref{fig:fit_2_models} that the LFUV scenario (red) can generate at most $C_{LL}^{\nu_{e}\nu_{e}}\simeq -0.7$ at the $1\sigma$ level while simultaneously fitting the rest of the $B$ meson anomalies, meaning a relative increment of $\sim10\%$ for $\mathrm{BR}(B^{+}\to K^{+}\nu\bar{\nu})$ with respect to the SM prediction\footnote{Non-minimal $Z^{\prime}$ models with vector leptoquarks and new vector-like fermions can generate a $50\%$ enhancement of $\mathrm{BR}(B^{+}\to K^{+}\nu\bar{\nu})$ at the one-loop level. See \cite{Fuentes-Martin:2020hvc} for more details.}. This deficit with respect to the Belle-II measurement is alleviated in our LFV scenario (blue), obtaining the required value for the Wilson coefficient $C_{LL}^{\nu_{e}\nu_{e}}\simeq -7$,  interfering constructively with the SM contribution. Assuming bi-large mixing in the PMNS matrix ($s_{13}\to 0$) and normal ordering for the neutrino parameters \cite{Esteban:2020cvm}, the best fit point for the LFV scenario is given by 

\begin{align}
M_{Z'} & =(5.2\pm0.9)\,\mathrm{TeV},\quad g_{L}^{bs}=-0.11\pm0.04,\quad\alpha_{L}=(229\pm9)\degree,\quad\epsilon=(21.0\pm5.0)\degree\nonumber \\
X_{R} & =0.1\pm0.1,\quad X_{L}^{e}=-0.01\pm0.08,\quad X_{L}^{\mu}=-2.9\pm0.1,\quad X_{L}^{\tau}=3.5\pm0.4,
\end{align}
with the parameter $\delta$ being unconstrained. The results for the inverted ordering scan are similar and only differ in allowing slightly smaller values for the lepton couplings.

\begin{figure}[ht]
\begin{centering}
\includegraphics[scale=0.35]{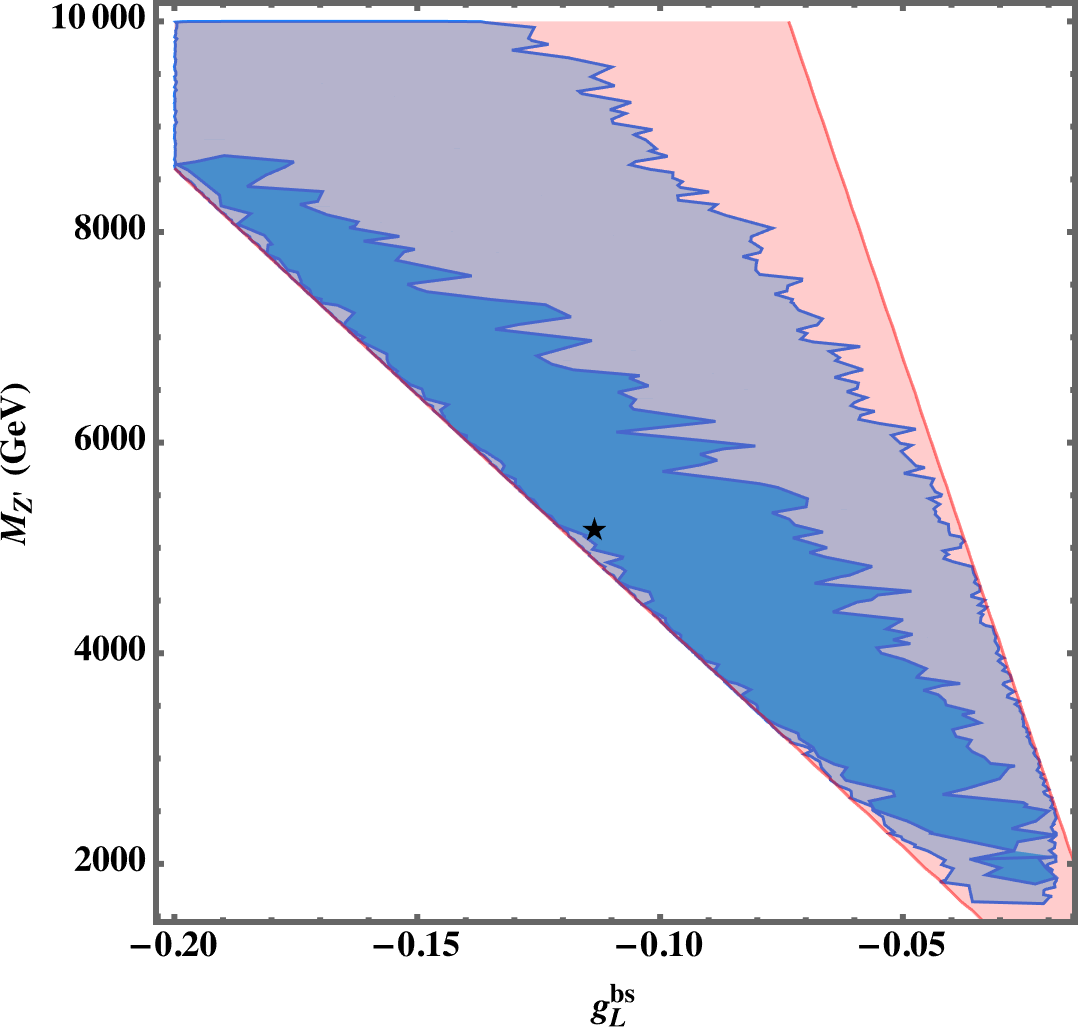} $\qquad$\includegraphics[scale=0.35]{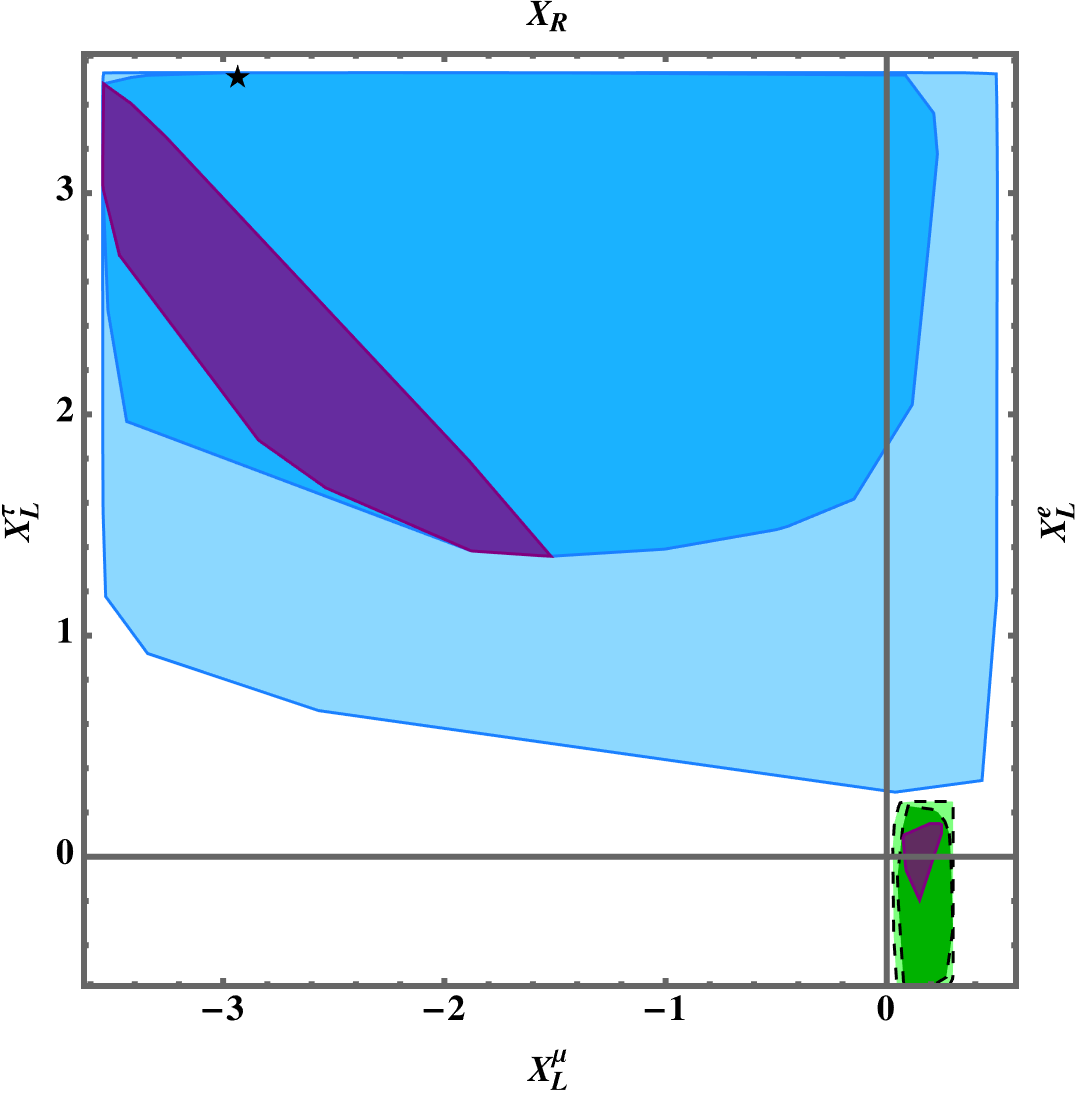} 
\par\end{centering}
\caption{Constraints to the parameter space of the LFV scenario fitting the $b\to s \ell^{+}\ell^{-}$ data simultaneously with $\mathrm{BR}(B^{+}\to K^{+}\nu\bar{\nu})$. Left: Projected $1$ and $2\sigma$ regions (blue and light blue respectively) in the $\{M_{Z'},\:g^{bs}_L\}$ plane constrained at $1\sigma$ level by $\Delta M_{s}$ with the region allowed by $\Delta M_{s}$ shown in red. Right:  The projected $1$ and $2\sigma$ regions constrained at $1\sigma$ level by $\Delta M_{s}$ is shown in the $\{X_{L}^{\mu},\,X_{L}^{\tau}\}$ plane using the same colours as the left panel and in the $\{X_{R},\,X_{L}^{e}\}$ using green and light green. The purple regions show how these regions shrink when we additionally include constrain our samples with $\mu\to e\gamma$ by post-processing the data.  We also include neutrino trident constraints but these do not have any impact on the contours in these planes.  \label{fig:constraints}}
\end{figure}

In Fig.\ref{fig:constraints} we consider additional constraints on the model. In both panels we additionally constrain our samples using the $\Delta M_{s}$ mass difference from $B_{s}-\overline{B}_{s}$ mixing \cite{DiLuzio:2019jyq,Alguero:2022est,Allanach:2022iod},
\begin{equation}
\left(\frac{g_{L}^{bs}}{0.52}\right)^{2}\left(\frac{10\mathrm{TeV}}{M_{Z'}}\right)^{2}=0.110\pm0.090.
\end{equation} and show the remaining $1\sigma$ and $2\sigma$ regions in blue and light blue respectively, while the constraint itself is indicated by the red shading in the left panel, where we present the results in the plane of the gauge boson mass and the $g^{bs}_L$ coupling. 
In the right panel of Fig.\ref{fig:constraints} we show these results (including the $\Delta M_{s}$ constraint) in planes featuring the lepton charges. We also consider constraints from  from $\mu\to e\gamma$ using Eq.(\ref{eq:MEG_bound}) and from the neutrino trident production cross section given by \cite{Altmannshofer:2014pba},
\begin{equation}
\frac{\sigma_{\mathrm{SM+NP}}}{\sigma_{\mathrm{SM}}}=1+\frac{\sqrt{2}}{G_{F}}\frac{g_{L}^{\mu\mu}}{M_{Z'}^{2}}\frac{(1+4\sin^{2}\theta_{w})(g_{L}^{\mu\mu}+g_{R}^{\mu\mu})+(g_{L}^{\mu\mu}-g_{R}^{\mu\mu})}{1+(1+4\sin^{2}\theta_{w})^{2}},
\end{equation}
which we required to be within $\sigma_{\mathrm{exp}}/\sigma_{\mathrm{SM}}=0.83\pm0.18$ \cite{Alguero:2022est}.  However we find that the neutrino trident constraint has no impact on our samples, which is in agreement with \cite{Altmannshofer:2014pba}, where the neutrino trident production was found to be less restrictive for TeV masses.  On the other hand, the $\mu\to e\gamma$ constraint can eliminate many scenarios in the full parameter space, with a clear impact (purple regions on the right panel of Fig.\ref{fig:constraints}) on the projected $1\sigma$ contours in the  $\{X_{L}^{\mu},\,X_{L}^{\tau}\}$ and $\{X_{R},\,X_{L}^{e}\}$ planes.  Note that this is from requiring $\mu\to e\gamma$ is fulfilled at the $1\sigma$ level, whereas the effect would be very mild if we relaxed this to $2\sigma$ since our parameter ranges were chosen to avoid lepton flavour constraints at the $2\sigma$ level. 

In both plots in Fig.\ref{fig:constraints} we observe an allowed region for the $Z'$ mass in the 2-8 TeV range at the $1\sigma$ level, although requiring large values for the $X_{L}^{\mu,\tau}$ couplings at the best fit point, near the perturbativity limit of $\sqrt{4\pi}$ and within reach of the future Muon Collider (MuC) \cite{MuonCollider:2022xlm,Altmannshofer:2023uci}. In contrast, rather small values for both the left-handed electron coupling (compatible with zero at the $1\sigma$ level) and for the right-handed universal coupling $X_R$ are preferred by the fit. The values of these lepton couplings will in general depend in the NP scenario chosen from the model-independent fits. However our fit results could provide a useful tool for constraining the quantum numbers for the SM charged leptons, and subsequently, for finding the values for the charges of the exotic fermions needed in the cancellation of the chiral anomalies. Such an algorithm for constructing anomaly-free tree-level $Z'$ models directly from experimental data is left for a follow-up work.

\begin{figure}[ht]
\begin{centering}
\includegraphics[scale=0.35]{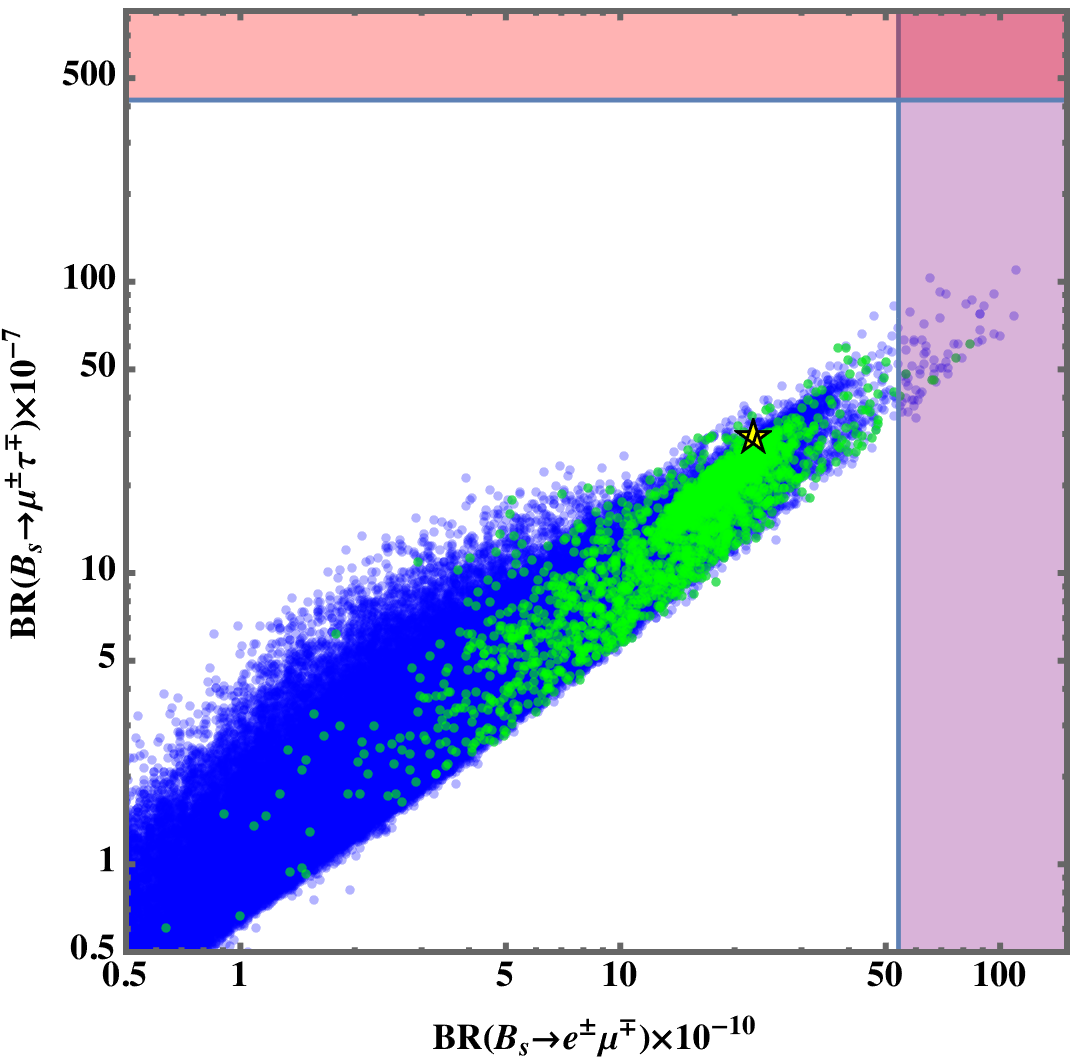} 
\par\end{centering}
\caption{Projected $1$ (green) and $2\sigma$ (blue) regions for the charged lepton flavour violating decays $\mathrm{BR}(B_s\to\mu^{\pm}\tau^{\mp})$ and $\mathrm{BR}(B_s\to e^{\pm}\mu^{\mp})$. The red and light purple regions are the current 90$\%$ C.L bounds taken from the Particle Data Group \cite{ParticleDataGroup:2022pth}.\label{fig:CLFV}}
\end{figure}

One of the main consequences from the LFV scenario is the prediction of the charged lepton flavour violating (CLFV) decays $B_s\to\ell^{\pm}\ell^{\prime\mp}$ and $B^+\to K^{+}\ell^{\pm}\ell^{\prime\mp}$. In particular, using the theoretical expressions from \cite{He:2021yoz}, in Fig.\ref{fig:CLFV} we show the generated values from the allowed parameter space for two of those CLFV modes, obtaining a good discovery potential associated to the $Z'$ lepton flavour violating scenario in the $B_s\to e^{\pm}\mu^{\mp}$ channel, within reach of future searches at both Belle II and the LHCb Upgrade II. 

\section{Summary and conclusions}\label{sec:conclusions}
We presented a model-independent update to some of the most common one and two dimensional scenarios fitting the neutral $B$ meson anomalies and corroborated that the $\{C_{9}^{\mu\mu},\:C_{10}^{\mu\mu}\}$ NP scenario is now disfavoured with respect to the scenarios fitted by $C_{9}^{\mu\mu}$ and $C_{9}^{ee}$. After this, and in view of the new Belle II measurement, we performed a 5 dimensional fit to both the $b\to s \ell\ell$ and $b\to s \nu\bar\nu$ observables and found a pull of $\textrm{Pull}_{\textrm{SM}}=5.9\sigma$. Based on this last result and using a quadratic approximation to the $\chi^2$ function, we then mapped out the resultant Wilson coefficients into the parameter space of two $U(1)'$ scenarios, one with LFUV and the second with LFV. We found that the LFUV scenario can generate an increment to the $B^+\to K^+\nu\bar{\nu}$ branching ratio of $10\%$, insufficient to explain the Belle II data. In the LFV scenario we showed that such branching ratio, including constraints from  $B_{s}-\overline{B}_{s}$ mixing and neutrino trident production, can be large enough once $\mathcal{O}(0.4)$ off-diagonal elements of the charged lepton rotation matrix were included. Finally, we presented predictions for CLFV decays, the most promising being the $B_s\to e^{\pm}\mu^{\mp}$ decay, within reach of future searches at both Belle II and the LHCb Upgrade II. 

\acknowledgments
We thank useful comments by B. C. Allanach regarding charged lepton flavour violating constraints. We also thank Lei Wu and Andrew Fowlie for discussions in the early stages of the project. The work of CS is supported by the Excellent Postdoctoral Program  of Jiangsu Province grant No. 2023ZB891. The work of PA is supported by the National Natural Science Foundation of China (NNSFC) under grant No. 12150610460 and by the supporting fund for foreign experts grant wgxz2022021L.


\bibliographystyle{JHEP}     

%
\bibliography{Belle_U1}

\end{document}